\documentclass[conference]{IEEEtran} 

\IEEEoverridecommandlockouts
\usepackage{cite}

\usepackage{amsmath,amssymb,amsfonts}
\usepackage{bbm}
\usepackage{graphicx}
\usepackage{textcomp}
\usepackage{xcolor}
\usepackage{algorithm}
\usepackage{algpseudocode}

\def\BibTeX{{\rm B\kern-.05em{\sc i\kern-.025em b}\kern-.08em
    T\kern-.1667em\lower.7ex\hbox{E}\kern-.125emX}}
\begin{document}

\title{Optimizing Intelligent Reflecting Surface-Base Station Association for Mobile Networks
}
\author{\IEEEauthorblockA{Dongzi~Jin\IEEEauthorrefmark{1}, Yong~Xiao\IEEEauthorrefmark{1}\IEEEauthorrefmark{4}, Yingyu Li\IEEEauthorrefmark{1}, Guangming~Shi\IEEEauthorrefmark{2}\IEEEauthorrefmark{4}, and Dusit Niyato\IEEEauthorrefmark{3} \\
		\IEEEauthorblockA{\IEEEauthorrefmark{1}School of Electronic Inform. \& Commun., Huazhong University of Science \& Technology, China}
		\IEEEauthorblockA{\IEEEauthorrefmark{2}School of Artificial Intelligence, Xidian University, Xi'an, China}
		\IEEEauthorblockA{\IEEEauthorrefmark{4}Pazhou Lab, Guangzhou, China}
		\IEEEauthorblockA{\IEEEauthorrefmark{3}School of Computer Science and Engineering, Nanyang Technological University, Singapore}
	}
}

\maketitle

\begin{abstract}
This paper studies a multi-Intelligent Reflecting Surfaces (IRSs)-assisted wireless network consisting of multiple base stations (BSs) serving a set of mobile users. We focus on the IRS-BS association problem in which multiple BSs compete with each other for controlling the phase shifts of a limited number of IRSs to maximize the long-term downlink data rate for the associated users. We propose MDLBI, a Multi-agent Deep Reinforcement Learning-based BS-IRS association scheme that optimizes the BS-IRS association as well as the phase-shift of each IRS when being associated with different BSs. MDLBI does not require information exchanging among BSs. Simulation results show that MDLBI achieves significant performance improvement and is scalable for large networking systems. 
\end{abstract}

\begin{IEEEkeywords}
Multi-IRS, Transmit Beamforming, Reflect Beamforming, BS-IRS association, Reinforcement Learning;
\end{IEEEkeywords}

\section{Introduction}
Intelligent Reflecting Surface (IRS) is a programmable meta-surface consisting of a large number of low-cost and passive reflecting elements. The phase shifts of these elements can be controlled and optimized to enhance the signal reception at the receiver. Recent studies suggest that the IRS has the potential to establish a reflected link with performance that is comparable to the Line-of-Sight (LoS) link without consuming any extra energy for signal power amplification\cite{scattermimo}. Due to its potential to improve the wireless communication performance with relatively low implementation and maintaining costs, IRS has been promoted by both industry and academia as the key enabling technology for next generation wireless communication systems\cite{gong2020survey}.

Recent report suggests that deploying multiple IRSs has the potential to further improve the signal reflection performance and alleviate the co-channel interference between different signal sources\cite{wu2020magzine}. However, enabling multiple IRSs to enhance the system performance introduces several challenges. First, it is known that, to maximize communication performance, an IRS needs to keep track of the channel state information (CSI) between itself and the signal source as well as the intended receivers. In the multi-IRS system, each IRS should not only coordinate with with its own associated sources and receivers but also carefully coordinate with other IRSs to avoid introducing the co-channel interference caused by reflecting signals toward unintended receivers. The total volume of coordination information among signal sources, IRSs and the receivers is expected to grow significantly with the number of IRSs and the number of elements of each IRS. Second, in a multi-user system consisting of multiple signal sources and receivers, how to improve the overall system performance by allocating different IRSs to serve different sources or receivers is still an open problem. This problem is further exacerbated by the fact that in mobile network system, users can constantly move from one location to another. In this case, dynamically evaluating and adjusting the BS-IRS association is critical for maximizing the long-term performance of IRS-assisted wireless system.

In this paper, we consider a mobile networking system consisting of multiple base stations (BSs), and each offers wireless services to users within an exclusive service area (e.g., cell). Multiple IRSs are deployed throughout the service areas of BSs to further enhance the downlink data communication performance from BSs to users. We consider a dynamic environment in which users can move between different service areas of BSs at different time. As mentioned earlier, optimizing the IRS-assisted data communication generally require constantly global coordination among all the BSs, IRSs, as well as the users. Inspired by recent success in learning-based methods\cite{walid2020icc,huang2020jsac,xiao1,xiao2}, we propose MDLBI, a multi-agent deep reinforcement learning-based BS-IRS association scheme to allow each BS to compete for IRSs to serve their users. In MDLBI, each BS can neither know the IRS selection policy of others nor communicate with other BSs. Each BS, however, can learn and maintain a parameterized actor function which maps its locally observed states into its optimal decision. MDLBI is easy to implement and scalable to large networking systems. Extensive simulation has been conducted. Our results show that MDLBI converges well compared to existing benchmark methods and can be directly applied into networks with a large number of BSs and IRSs. To the best of our knowledge, this is the first work to study the BS-IRS association problem in a dynamic networking environment.

The reminder of this paper is organized as follows. Section II presents the related work. System model and problem formulation are described in Section III. We present the proposed MDLBI in Section IV. The simulation results are presented in Section V and paper is concluded in Section VI.

\section{Related Work}
\textbf{Single IRS-assited wireless network:}
Most existing works on IRS-assisted wireless networks focus on optimizing the single-IRS scenarios. Particularly, in\cite{wu2019twc}, the authors applied IRS to minimized the total transmit power at the BS. A joint optimization problem of both transmit beamforming of active BS antennas and the reflect beamforming of passive phase shifters has been investigated. The authors in \cite{walid2020icc} applied deep reinforcement learning to optimize the overall energy efficiency for IRS powered by harvested energy. The authors in \cite{huang2020jsac} adopted a deep reinforcement learning method to optimize the transmit beamforming and reflect beamforming in dynamic environment. The authors in\cite{tan2016icc} proposed a novel phase shift solution for IRS to minimize the co-channel interference for multiple receivers sharing the same spectrum. Recently, IRS is also utilized to maximize the secrecy rate of the legitimate communication link \cite{zhangrui2019letters} and extend the wireless coverage with ultra-reliable low-latency communication services \cite{yang2020letters}.

\textbf{Multi-IRS-assisted wireless network:} Multi-IRS-assisted network has attracted significant interest due to its potential to significantly improve the spectrum and energy efficiency\cite{sun2020wcnc,yu2020jsac,wu2020jsac,hu2020tcom,weidong}. For example, the authors in \cite{sun2020wcnc} studied the joint optimization problem of the transmit beamforming, reflect beamforming, and the set of active IRSs to minimize both the transmit power consumption of the BS and circuit power consumption of the IRSs. The authors in\cite{yu2020jsac} investigated the joint design of transmit beamforming and artificial noise covariance matrix at an access point and the reflect beamforming at the IRSs to maximize the system sum rate while limiting the maximum information leakage at the potential eavesdroppers. The joint active and passive beamforming optimization problem for IRS-assisted simultaneous wireless information and power transfer (SWIPT) is studied in \cite{wu2020jsac}. Different from the full knowledge channel state information (CIS) assumption  in \cite{sun2020wcnc,yu2020jsac,wu2020jsac}, the authors in \cite{hu2020tcom} designed a transmit and reflect beamforming solution based on the imperfect location information of users. In \cite{weidong}, the authors performs an initial investigation on the IRS association problem, where the dynamics of environment is not considered.

\section{System Model And Problem Formulation}

\subsection{System Model}

We consider a multiple IRS-assisted network consisting of $M $ BSs, each has $N_b $ antennas, that provides services to $K$ single-antenna users in the considered area, as shown in Fig. 1. Let $\mathbb{M}=\{b_1, b_2,\ldots,b_M\}$ and $\mathbb{K}=\{1,2,\ldots,K\}$ be the sets of BSs and users, respectively. Each BS covers an exclusive sub-region in the service area. We consider a mobile network in which users can move from one sub-region to another. The user mobility can be regarded as a slotted process in which the set of users located in the sub-region of each BS can be considered as fixed within each time slot, i.e., we use {$\mathbb{C}_{m,t}$} to denote as the set of users served by BS $b_m$ during time slot $t$ and $\cup_{m \in \mathbb{M}}\mathbb{C}_{m,t}=\mathbb{K}$. To simplify our description, we focus on the downlink communication and the main objective of each BS is to maximize the data rates from itself to the users. We assume a proper inter-cell interference cancellation mechanism has been adopted between BSs and thus the data transmission of each BS does not cause any noticeable interference to the users located in the coverage area of other BSs.

Suppose $L$ IRSs are deployed in the service area that can be utilized by BSs to improve the downlink data communication performances of BSs. Let $\mathbb{L}=\{1, 2,\ldots, L\}$ be the set of all the IRSs and $N_l$ be the number of passive reflecting elements of the $l$th IRS for $ l \in \mathbb{L}$. BSs compete for the control of IRSs to serve their users. We assume each IRS can only serve a single BS in each time slot. Each BS, however, can take control of multiple IRSs. $\mathbb{N}_{m,t}$ denotes the IRS set controlled by BS  $b_m$ during time slot $t$ and we have  $\cup_{m \in \mathbb{M}}\mathbb{N}_{m,t}=\mathbb{L}$.

\begin{figure}[htbp]
	\centering
	\includegraphics[width=7.5cm]{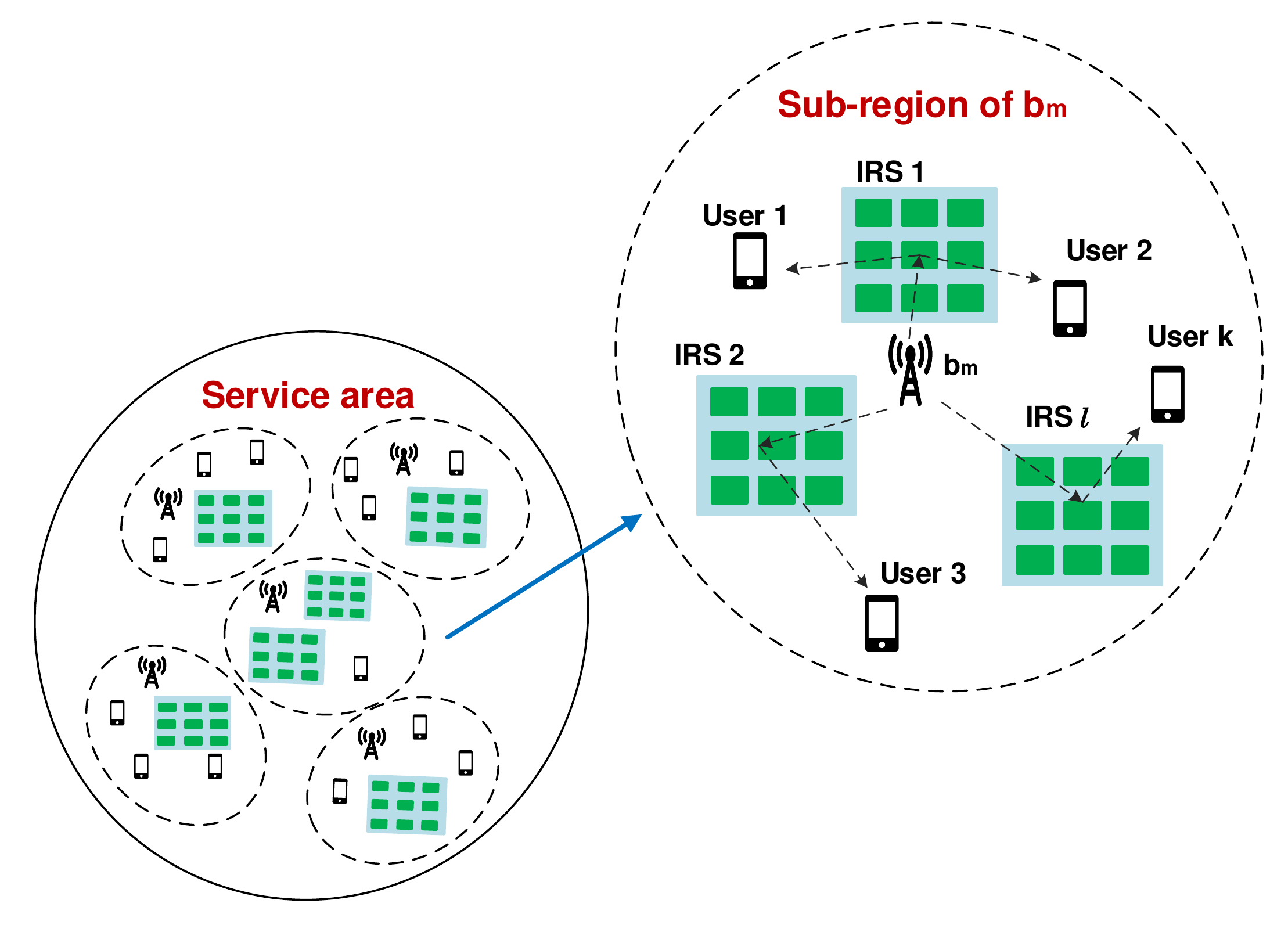}
	\caption{System Model.}
	\label{fig1}
\end{figure}

Let $\mathbf{G}_{m, l} \in \mathbb{C}^{N_l \times N_{b}}$ and $\mathbf{h}_{l, k}^{H} \in \mathbb{C}^{1 \times N_l}$ to be the complex equivalent baseband channel vectors between each BS $b_m$ and the $l $th IRS, and between the $l $th IRS and the $k $th user, respectively. The channel gain between BS $b_m$ and the $k$th user is given by $\mathbf{H}_{m,k} \in \mathbb{C}^{1 \times N_b}$. We assume  $\mathbf{H}_{m,k}$, $\mathbf{G}_{m, l}$, and $\mathbf{h}_{l, k}^{H}$ can be regarded as constants during each transmission time slot\cite{sun2020wcnc}. Furthermore, the channel state information (CSI) is perfectly available at the BSs as assumed in\cite{sun2020wcnc,huang2020jsac,yu2020jsac,wu2019twc}. Let $\mathbf{\Phi}_{l}=\operatorname{diag}\left\{\left[ e^{j \theta_{l}^{1}}, \ldots, e^{j \theta_{l}^{N_l}}\right]\right\}$ be the phase shift matrix of the $l $th IRS, and $\theta_{l}^{e} \in [0, 2\pi]$ be the $e $th phase shift of the $l $th IRS, where $\operatorname{diag}\{\cdot\}$ denotes diagonal matrix and $e \in \{1, \ldots, N_l\}$. We write $\mathbf{\Phi}=\operatorname{blkdiag}\left(\mathbf{\Phi}_{1}, \mathbf{\Phi}_{2}, \ldots, \mathbf{\Phi}_{l} \right) $ as the phase shift matrices of all IRSs, where $\operatorname{blkdiag}\{\cdot\}$ denotes block diagonal matrix. Let $n_{k} \sim \mathcal{C} \mathcal{N}\left(0, \sigma_{k}^{2}\right)$ be the additive white Gaussian noise where $\sigma_{k}^{2} $ is the received noise power of user $k$. 

Let $\mathbf{A}=\operatorname{blkdiag}\left[\mathbf{A}_{1}, \mathbf{A}_{2}, \ldots,\mathbf{A}_{M}\right] $ be the BS-IRS association matrix, where $\mathbf{A}_{m} \in \mathbb{C}^{K \times L} $ is the BS-IRS association matrix of the $m $th BS. $\left[\mathbf{A}_{m}\right]_{k^\prime,l^\prime}$ is a binary variable specifying the association relationship among the $m$th BS, the $k^\prime $th user, and the $l^\prime $th IRS. In other words, $\left[\mathbf{A}_{m}\right]_{k^\prime, l^\prime} =1 $ means that the $m $th BS $b_m$ allocates the $l^\prime $th IRS to assist the downlink communication to the $k^\prime$th user, where $l^\prime \in \mathbb{N}_m$ and $k^\prime \in \mathbb{C}_m$. When focusing on a specific time slot $t$, we write $\mathbb{N}_{m,t} $ and $\mathbb{C}_{m,t}$ as $\mathbb{N}_m $ and $\mathbb{C}_m$ for simplification.  Since each user is assumed to be served by a single IRS, we have $\sum_{l^\prime \in \mathbb{N}_m}\left[\mathbf{A}_{m}\right]_{k^\prime, l^\prime}=1, \forall k^\prime \in \mathbb{C}_m$. We can use $\mathbf{\Phi}_{k^\prime}=\sum_{l^\prime \in \mathcal{L}} \mathbf{\Phi}_l^\prime \mathbbm{1}_{\{\left[\mathbf{A}_{m}\right]_{k^\prime, l^\prime}=1\}}$ to represent the phase shift matrix of the single IRS allocated to the ${k^\prime}$th user by the $m$th BS $b_m$, where $\mathbbm{1}_{\{.\}}$ is the indicator function.

The received signal at the $k^\prime$th user served by BS $b_m$ can be written as
\begin{equation}
	\begin{aligned}
	y_{m,k^\prime} =\mathbf{H}_{m,k^\prime}\mathbf{x}+\mathbf{h}_{l, {k^\prime}}^{H} \mathbf{\Phi}_{k^\prime} \mathbf{G}_{m, l} \mathbf{x}+n_{k^\prime},
    \end{aligned}
\end{equation}
where $\left(.\right)^H$ is conjugate transpose.

The transmitted signal of the BS $b_m$ can be expressed as $\mathbf{x}=\sum_{{k^\prime} \in \mathbb{N}_m} \mathbf{w}_{m,k^\prime} s_{k^\prime}$, where $s_{k^\prime} $ denotes the desired signal of the ${k^\prime} $th user with $s_{k^\prime} \sim \mathcal{C} \mathcal{N}(0,1)$, and $\mathbf{w}_{m,k^\prime} \in \mathbb{C}^{N_{b} \times 1}$ is the transmit beamforming vector for the $k^\prime $th user. Each BS has a maximum transmit power constraint: 
\begin{equation}
\sum_{{k^\prime} \in \mathbb{C}_m} \left\|\mathbf{w}_{m,k^\prime}\right\|^{2} \leq P_{\max }, \forall m \in \mathbb{M} . 
\end{equation}
By substituting $\mathbf{x}$ into (1), we have
\begin{equation}
	y_{m,{k^\prime}} = \mathbf{H}_{m,{k^\prime}}\sum_{i \in \mathbb{N}_m} \mathbf{w}_{m,i} s_{i}+\mathbf{h}_{l, {k^\prime}}^{H} \mathbf{\Phi}_{k^\prime} \mathbf{G}_{m, l} \sum_{i \in \mathbb{N}_m} \mathbf{w}_{m,i} s_{i}+n_{k^\prime}.	
\end{equation}

Accordingly, the SINR at the $k $th user served by BS $b_m$ is given by
\begin{equation}
	r_{m,{k^\prime}}=\log \left(1+\frac{\left|\mathbf{H}_{m,{k^\prime}}\mathbf{w}_{k^\prime}+\mathbf{h}_{l, {k^\prime}}^{H} \mathbf{\Phi}_{k^\prime} \mathbf{G}_{m, {k^\prime}} \mathbf{w}_{k^\prime}\right|^{2}}{\sum_{i \neq {k^\prime}}^{K}\left|\mathbf{h}_{l, {k^\prime}}^{H} \mathbf{\Phi}_{k^\prime} \mathbf{G}_{m, {k^\prime}} \sum_{i \in \mathbb{N}_m} \mathbf{w}_{i}\right|^{2}+\sigma_{k^\prime}^{2}}\right).
\end{equation}

The achievable sum rate of a multiple IRS-assisted wireless communication system is given by
\begin{equation}
	R=\sum_{m \in \mathcal{M}}\sum_{k^\prime \in \mathbb{C}_m} r_{m,{k^\prime}}.
\end{equation}

\subsection{Problem Formulation}
Given the defined system model, our goal is to maximize the sum rate of the multiple IRS-assisted communication system by jointly optimizing the user scheduling and association matrix $\mathbf{A} $, phase shifts matrix $\mathbf{\Phi} $ and transmit beamforming matrix $\mathbf{w} $, i.e., the joint optimizing problem can be written as follows:
\begin{subequations}
	\begin{align}
		&(\mathcal{P}):  \max _{\mathbf{w}, \mathbf{\Phi}, \mathbf{A}} R, \\
	     \mbox{s.t.} & \sum_{k^\prime \in \mathbb{C}_m} \left\|\mathbf{w}_{m,{k^\prime}}\right\|^{2} \leq P_{\max } ,\forall m \in \mathcal{M}, \\
		& \left|e^{\theta_{l^\prime}^{m}}\right|=1, \forall m \in \mathcal{M}, l^\prime \in \mathbb{N}_m, \\
		& \left[\mathbf{A}_{m}\right]_{k^\prime, l^\prime} \in \{0,1\}, \forall m \in \mathcal{M}, l^\prime \in \mathbb{N}_m, k^\prime \in \mathbb{C}_m.
	\end{align}
\end{subequations}

We can observe that the objective function of problem $(\mathcal{P}) $ is generally non concave and the three optimization variables $\mathbf{w} $, $\mathbf{\Phi} $, and $\mathbf{A} $ are coupled with each other which make the problem difficult to solve. Besides, the constraint (6c) is highly non-convex as the phase of each element is forced to have a unit magnitude. In the rest of the paper, we propose a Multi-agent Deep Deterministic Policy Gradient (MDDPG)-based solution, called MDLBI to jointly optimize transmit beamforming $\mathbf{w} $,  reflect  beamforming $\mathbf{\Phi} $, and BS-IRS association $\mathbf{A} $.

\section{Multi-agent DDPG for the Multiple IRSs-assisted Communication System }
In this section, we introduce a MDDPG-based solution, called MDLBI for optimizing the multiple IRSs-assisted communication system.   

In this method, each BS can observe the state including the channel information  (i.e., $\mathbf{H}_{m,k^\prime}$, $\mathbf{G}_{m, l^\prime}$, $\mathbf{h}_{l^\prime, k^\prime}^{H} $, and leakaged control signal, $ \forall l^\prime \in \mathbb{N}_m, \forall k^\prime \in \mathbb{C}_m$) , output the actions (i.e., phase shifts $\mathbf{\Phi}_{l^\prime}, \forall l^\prime \in \mathbb{N}_m $, the transmit beamforming $\mathbf{w}_m, \forall m \in \mathbb{M}$, and the association matrix $\mathbf{A}_m, \forall m \in \mathbb{M}$), and obtain reward (i.e., sum rate $\sum_{k^\prime \in \mathbb{C}_m}r_{k^\prime} $) during each time slot $t$, as shown in Fig. 2. 

\begin{figure}[htbp]
	\centering
	\includegraphics[width=7cm]{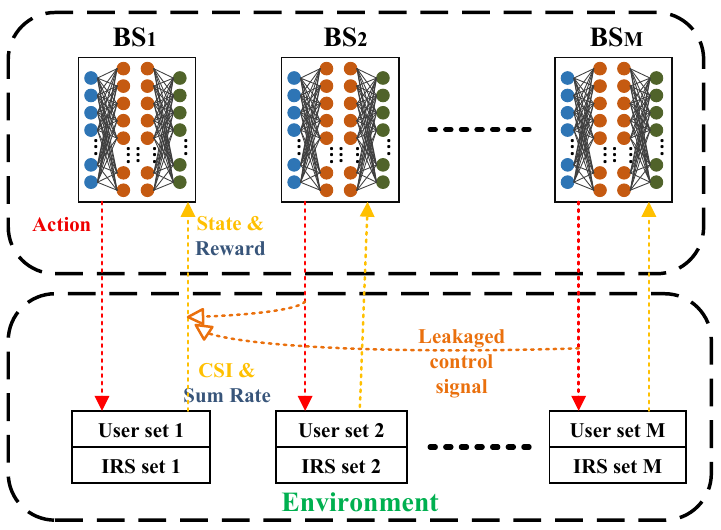}
	\caption{MDDPG based optimization for the multiple IRSs-assisted communication system}
	\label{fig2}
\end{figure}

One of the main advantages for adopting MDDPG-based method to address problem $(\mathcal{P}) $ is that MDDPG is applicable to continuous and high-dimensional action spaces. This makes it suitable for our problem, in which the phase shifts $\mathbf{\Phi} $ are continuous and the association matrix $\mathbf{A} $ is high-dimensional.

DDPG is a deep reinforcement learning algorithm that could operate over continuous action space by maintaining an actor network to specify the current policy deterministically by mapping the observed state to a specific action. Besides, DDPG requires fewer steps of experiences than Deep Q Network (DQN) to find optimal solutions in the Atari domain\cite{ddpg}. Under the actor-critic architecture, DDPG maintains two networks,  namely the actor network and the critic network. The critic network is trained to approximate the Q-table using neural networks without the curse of dimension, while the actor network is trained to generate a deterministic policy instead of policy gradient. Furthermore, target networks are also adopted to improve the stability\cite{dqn_nature}. However, DDPG is not specialized for the multi-agent environment since the environment is non-stationary from the perspective of each agent (i.e. BS). By utilizing of the control signal between the BSs and IRSs, we extend DDPG into a multi-agent version for our task. More specifically, there are four sub-nets, namely the critic network $Q_{\theta_{m}^\prime}^\mu (o_m, a_m) $, the target critic network $Q_{\theta_c}^{\mu^\prime} (o_m, a_m) $, the actor network $\mu_{\theta_m} (o_m)$ (abbreviated as $\mu_{m}$), and the target actor network $\mu_{\theta^\prime}^\prime (o_m)$. We use $\theta_{m^\prime}$, $\theta_m$, $\theta_c$, and $\theta^\prime$ to denote the parameters of the critic network, the actor network, and the target networks during certain time slots, respectively. Note that  $a_m$ is the action of $b_m$ at the current time slot, and $o_m$ is the current state of BS $b_m$. It contains two parts $o_m=(s_m, \rho_{-m})$, where $s_m$ is the local state of physical environments (i.e., all channel state information in the sub-region of $b_m$ ) and $\rho_{-m}$ represents the other BS agents' strategies to dominate the IRS sets. This observation of other BSs' strategies to choose certain IRS sets can be realized by capturing the leakage of control siganls between other BSs and IRSs, since the control of IRS are typically connected with the BS through wireless\cite{gong2020survey}.

With the above notations, we formally describe the construction of the model in detail as follows.

\noindent
\textbf{State Space:} $o_m \in \mathbb{O}_m$ is the state of $b_m$, which is determined by the physical environment (direct channel $\mathbf{H}_{m,k}$, channel between $b_m$ and the $l$th IRS $\mathbf{G}_{m, l}$ and channel between the $l $th IRS and the $k $th user $\mathbf{h}_{l, k}^{H},  \forall l \in \mathbb{N}_m, \forall k \in \mathbb{C}_m$) and the other BS agents' strategies $\rho_{-m}$. Since neural network can only take real numbers and the channel information are complex, we separate the real and imaginary parts of the channel information as independent inputs. There are  $2card(\mathbb{C}_m)N_b$, $2card(\mathbb{N}_m)N_bN_l$, $2card(\mathbb{N}_m)N_lcard(\mathbb{C}_m)$ and $\left(M-1\right)L$ entries respectively contributed by $\mathbf{H}_{m,k}$, $\mathbf{G}_{m, l}$,  $\mathbf{h}_{l, k}^{H}$ and $\rho_{-m}$, where $card(\cdot)$ represents the cardinality of a set. Hence, the total number of entries for sate action is $D_s=2card(\mathbb{C}_m)N_b$+ $2card(\mathbb{N}_m)N_bN_l+2card(\mathbb{N}_m)N_lcard(\mathbb{C}_m)+\left(M-1\right)L$.

\noindent
\textbf{Action Space:} $a_m \in \mathbb{A}_m$ is the action at current state, which is constructed by  phase shifts $\mathbf{\Phi}_l, \forall l \in \mathbb{N}_m $, transmit beamforming vectors $\mathbf{w}_k, \forall k \in \mathbb{C}_m$ and association matrix $\mathbf{A}_m$. Likewise, there are $2N_bcard(\mathbb{C}_m)$, $card(\mathbb{N}_m)N_l $ and $L$ entries of action contributed by  $\mathbf{w}_k, \forall k \in \mathbb{C}_m $, $\mathbf{\Phi}_l, \forall l \in \mathbb{N}_m$ and $\mathbf{A}_m$, respectively. Hence, the total number of entries for action space is $D_a=2N_bcard(\mathbb{C}_m)+card(\mathbb{N}_m)N_l+L$.

\noindent
\textbf{Reward:} $r_m$ represents the instant reward defined as the sum rate of users in the sub-region of $b_m$, which can be obtained by knowing the other BS agents' strategies $\rho_{-m}$, the instantaneous channel information $\mathbf{H}_{m,k},\forall k$, $\mathbf{G}_{m, l} \forall l$ and  $\mathbf{h}_{l, k^\prime}^{H} \forall {l,k^\prime}$ and the action (i.e.,  $\mathbf{w}_{m,k^\prime}, \forall k^\prime $, $\mathbf{\Phi}_l, \forall l \in \mathbb{N}_m $ and $\mathbf{A}_m$) obtained from the actor network.

The expected reward of each BS $b_m$ is given by
\begin{equation}
	J\left(\theta_{m}\right)=\mathbb{E}_{s \sim p^{\mu}, a_{m} \sim {\mu}_{m}}\left[\nabla_{\theta_{m}} \log {\mu}_{m}\left(a_{m} \mid o_{m}\right) Q_{m}^{\mu}\left(o_m,a_m\right)\right],
\end{equation}
where $p^{\mu}$ is the state distribution.


%

Considering the continuous of action space, the gradient for the parameter $\theta_m$ of the deterministic policy $\mu_{\theta_m}$ (abbreviated $\mu_{m}$) is given as 
\begin{equation}
\begin{aligned}
	\nabla_{\theta_{m}} J\left(\boldsymbol{\mu}_{m}\right)=& \mathbb{E}_{o \sim \mathcal{D}}\left[\nabla_{\theta_{m}} \boldsymbol{\mu}_{m}\left(a_{m} \mid o_{m}\right)\right.\\
	&\left.\left.\cdot \nabla_{a_{m}} Q_{m}^{\mu}\left(o_m\right)\right|_{a_{m}=\boldsymbol{\mu}_{m}\left(o_{m}\right)}\right]
\end{aligned}	
\end{equation}

where $\mathcal{D}$ is the experience buffer contain trumples $\left( o_m, o_m^\prime, r_m \right)$, recording the experiences of all the agent BS.

Then the critic network is updated by minimizing the following loss function:
\begin{equation}
	\mathcal{L}\left(\theta_{m}^\prime \right)=\mathbb{E}_{o_m,o_m^\prime, r_m}\left[\left(Q_{m}^{\mu}\left(o_m\right)-y\right)^{2}\right],
\end{equation}

\begin{equation}
	y=r_{m}+\left.\gamma Q_{m}^{\boldsymbol{\mu}^{\prime}}\left(o_m\right)\right|_{\rho_{-m}},
\end{equation}
where $\mu^\prime$ is the set of target policies.

Finally, the agent softly updates the target networks with a small instant $\tau \ll 1$, i.e.,
\begin{equation}
	\begin{array}{l}
		\theta^{\prime} \leftarrow \tau \theta_{m}+(1-\tau) \theta^{\prime}.
	\end{array}
\end{equation}
\begin{equation}
	\begin{array}{l}
		\theta_c \leftarrow \tau \theta_{m}^\prime+(1-\tau) \theta_c.
	\end{array}
\end{equation}
This means the target networks are changed in a much slower speed than the actor and critic networks, which greatly improving the stability of the learning\cite{ddpg}. 


Given the set of IRS $\mathbb{N}_m$, the BS $b_m$ interacts with the environment in a trial-and-error manner to optimize the sum rate of user set $\mathbb{C}_m$ in its sub-region. During each time step $t $ of an episode, each BS (e.g. $b_m$) observes the current state $o_m$, applies a action $a_m $ defined by policy $\mu_{\theta_m} $ to the environment, and obtains the instant reward $r_m$. 

The details of the proposed algorithm are presented in Algorithm 1. \textbf{At the beginning of the algorithm,} parameters of the BSs and environment are initialized. In this paper, the experience buffer $\mathcal{D}$, the other users' trategies $\rho_{-m}$, the actor network parameters $\theta_{m}$,  the target network parameters $\theta^\prime$, the critic network parameters $\theta_{m}^\prime$ and the association matrix $\mathbf{A}$ are randomly initialized. And the transmit beamforming $\mathbf{w}$, the phase shifts  $\mathbf{\Phi}$ are simply initialized as identity matrix. \textbf{After initialization}, the new experience $(s_t, a_t, r_{t+1}, s_{t+1}) $ are collected into the experience buffer $\mathcal{B}$ (i.e., the step 4-8). A minibatch of experience with size $ w$ is randomly sampled from $\mathcal{B}$, i.e., step 9. The step 10 and 11 describe the update of the critic network and the actor network. Finally, the target networks are updated, i.e., the step 12. The algorithm run over $N$ episodes and each episode with $T$ time steps. During each episode, the algorithm terminates whenever it converges or finishes the $T$ time steps. 

	\begin{algorithm}[h!]  
	\caption{Multi-agent DDPG based optimization with IRS-BS association $b_m$} 
	
	{\bf Output}: {$a_m=\{\mathbf{w}_m, \mathbf{\Phi}_l \forall l \in \mathbb{N}_m, \mathbf{A}_k\}$}, {$Q$} value function
	
	{\bf Initialization}:  experience buffer {$\mathcal{D}$} with size {$D_l$}, the actor network parameter {$\theta_{m}$}, the target networks parameter {$\theta^\prime$}, the critic network parameter {$\theta_{m}^\prime$}, the transmit beamforming {$\mathbf{w}_m$}, the phase shifts {$\mathbf{\Phi}_l, \forall l \in \mathbb{N}_m$} and the association matrix {$\mathbf{A}_m$};
	\begin{algorithmic}[1] 
		
		\For{$episode = 0,1,2, \ldots, N-1$}

		\State Collect the channel information $\mathbf{H}_{m,k},\forall k \in \mathbb{C}_m$, $\mathbf{G}_{m, l}, \forall l \in \mathbb{N}_m$ and  $\mathbf{h}_{l, k}^{H}, \forall l \in \mathbb{N}_m, \forall k \in \mathbb{C}_m$ for the {$n$}th episode to obtain the first state {$s_0$};
		\For{$t=0,1,2, \ldots, T-1$}
		\State Obtain the IRS set $\mathbb{C}_{m,t}$ dominate by the BS $b_m$
		\State Observe action {$a_m=\{\mathbf{w}_m, \mathbf{\Phi}_l \forall l \in \mathbb{N}_{m,t}, \mathbf{A}_k\}$}= $\mu(\theta_m \mid o_m) $ from the actor network;
		\State Observe the next state $o_{m}^\prime$ given action $a_m$;
		\State Obtain the reward $r_m$;
		\State Store the experience $\left( o_m, a_m, r_m, o_{m}^\prime \right)$ in the 
		experience buffer $\mathcal{D}$;
		\State Sample a random minibatch of $W$ transitions from 
		the experience buffer $\mathcal{D}$;
		\State Update the critic network by minimizing the loss 
		as described in Eq. (9);
		\State Update the actor network using the policy gradient 
		as described in Eq. (8);
		\State Update the target networks using Eq. (11) and (12);
		\EndFor		
		\EndFor  			    
	\end{algorithmic}  
\end{algorithm}

\section{Stimulation and Results}
In this section, we evaluate the performance of the proposed MDDPG-based algorithm. All the channels are assumed to suffer from both path loss and Rayleigh fading. As in \cite{sun2020wcnc,huang2019twc}, the path loss exponents of the BS-IRS channel and the IRS-user channel are set as 2.5 and 2.4, respectively. The actor and critic network are both fully connected deep neural network with one input layer, two hidden layers and an output layer. The output layer of the actor network has the same dimension as the action and we use tanh function as the activation function. The output of the critic network is a scalar with one dimension. The activation function of all the hidden layers is Relu function.

\begin{table} [th]
	\centering
	\footnotesize
	\caption{Simulation Congiguration} 
	\setlength{\tabcolsep}{3mm}{
		\renewcommand\arraystretch{1.5}
		\begin{tabular} {|c|c|c|} 
			\hline
			Parameters & Description & Value\\ 
			\hline 
			$\gamma$ & Discount factor & 0.99 \\
			\hline
			$\mu_{c}$ & Learning rate of the actor & 0.0001\\
			\hline
			$\mu_{a}$ & Learning rate of the critic & 0.001 \\
			\hline
			$\tau$ & Soft target update factor & 0.001 \\
			\hline
			$D$ & Experience buffer Size & 10000 \\
			\hline
			$N$ & Number of episodes & 5000 \\
			\hline
			$T$ & Maximum time steps of each episode & 800 \\
			\hline
			$W$ & Batch size & 64 \\
			\hline
	\end{tabular}}
	\label{tab usecase}
\end{table}

In Fig.3, we investigate the convergence performance of MDLBI compared to the original DDPG-based solution. Since the batch sampling is utilized, we use time as the $x $ axis instead of time steps. We can observe that the proposed solution converges much faster and can obtain a  higher sum rate at around 8.5 bps/Hz. This means that the proposed MDLBI could improve the transmit beamforming compared to DDPG-based method. In practical system, each BS could utilize the leakage control signals of IRSs sent by other BSs to estimate their competition over the IRSs. In this way, the possibility of collisions between BSs when competing for the same IRS can be reduced.
\begin{figure}[htbp]
	\centering
	\includegraphics[width=8cm]{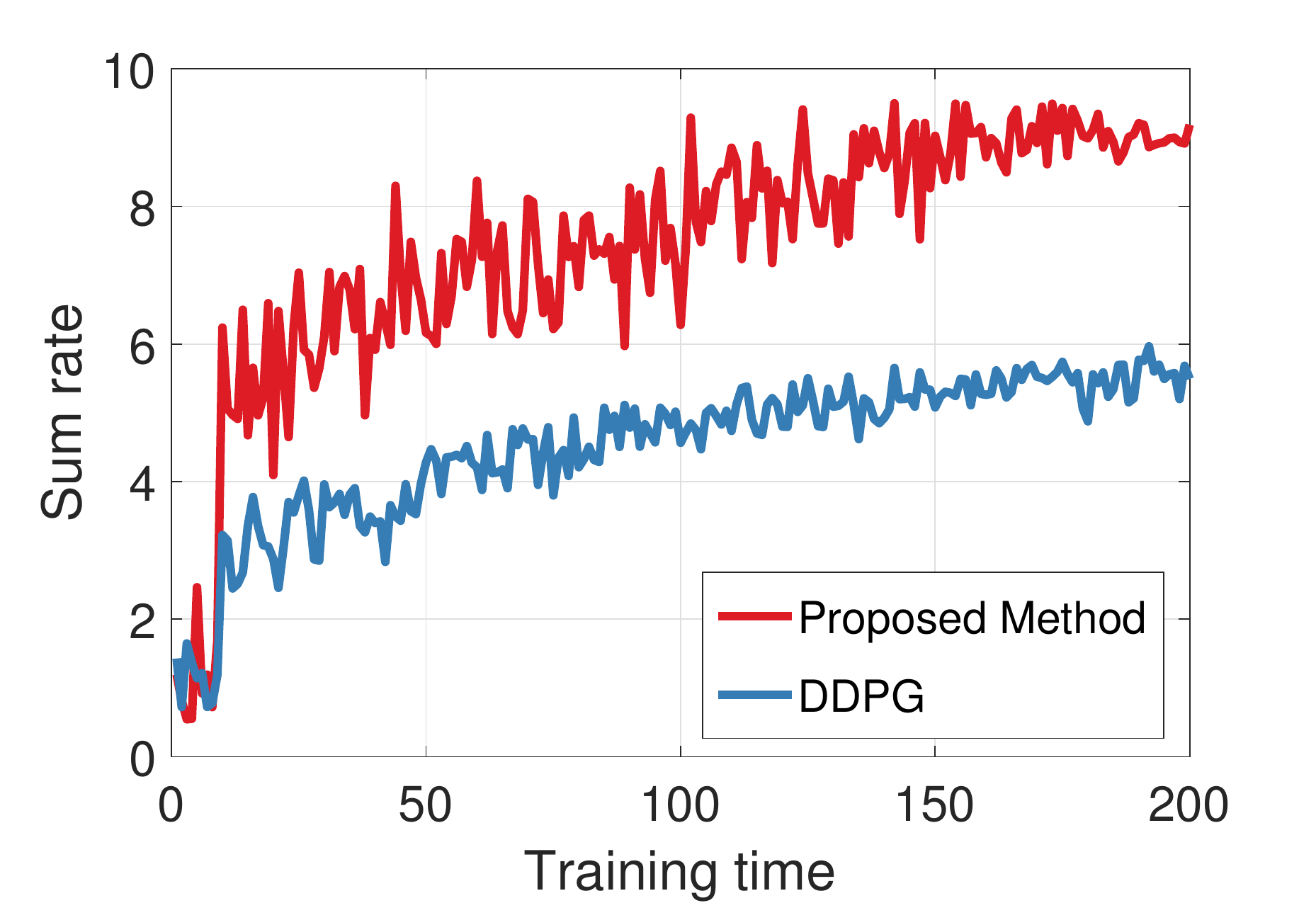}
	\caption{Instant reward (dB) as a function of time}
	\label{fig3}
\end{figure}

In Fig.4, we consider the instant reward (i.e. sum rate) under different system settings $L=\{1,2,4\}$,$ K=\{10,30\} $. The number of BS $M$ is fixed to 2. It can be observed that the sum rate of the service area increases with the number of IRSs. By comparing cases with $\{K=10,L=1\}$ and $\{K=30,L=1\}$, we can observe that the performance enhancement induced by a single IRS increased a little with the number of served users. However, when more IRSs can be deployed in the service area, such as cased with $\{K=30,L=2\}$ and $\{K=30,L=4\}$, the sum rate could be further improved. This result verifies that deploying  more number of IRSs even in a distributed manner could improve the system capacity. 

Fig. 5 compares the sum rate of our proposed method to that of a fixed BS-IRS association solution. It can be observed that although the optimization with a fixed association strategy may obtain higher sum rate at the beginning of the time of consideration, the overall system performance may fluctuate with time due to the user mobility.

\begin{figure}[htbp]
	\centering
	\includegraphics[width=9.5cm]{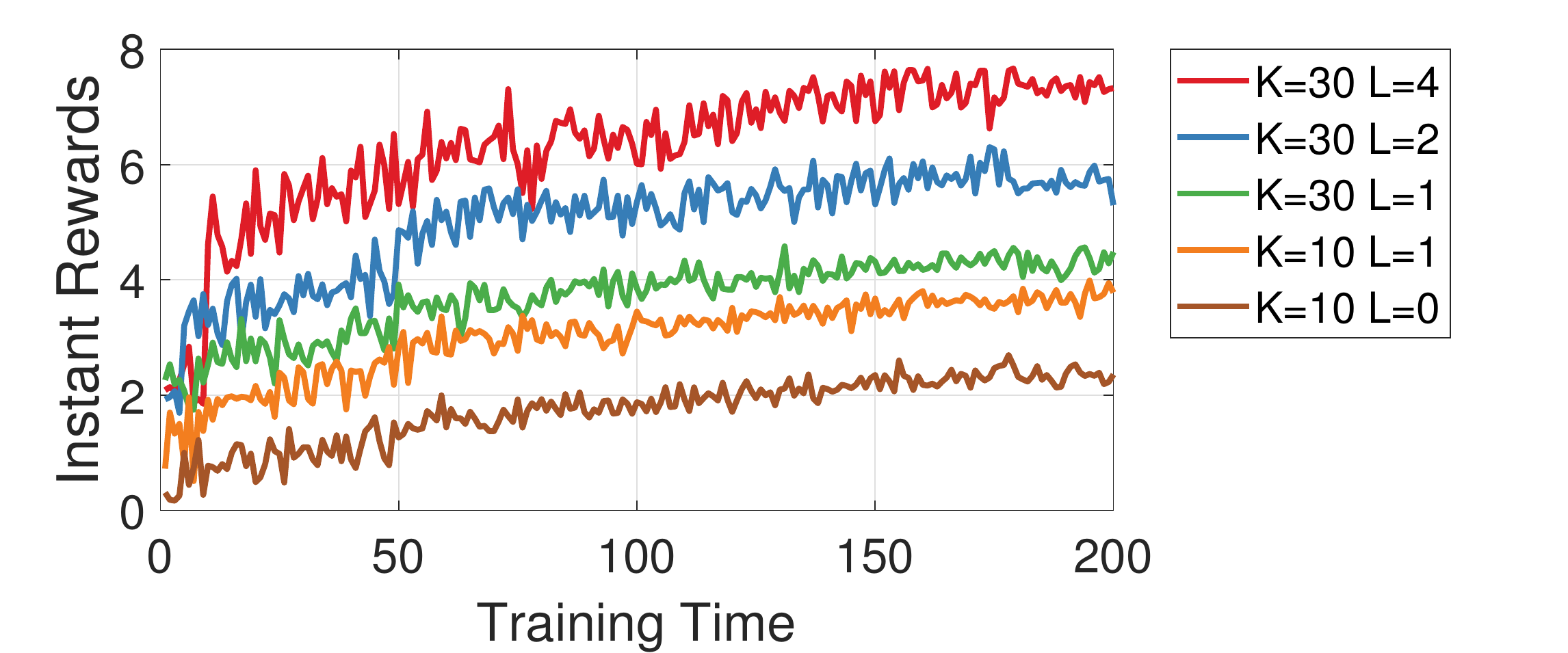}
	\caption{Instant reward (dB) as a function of time under different system settings}
	\label{fig4}
\end{figure}

\begin{figure}[htbp]
	\centering
	\includegraphics[width=9.5cm]{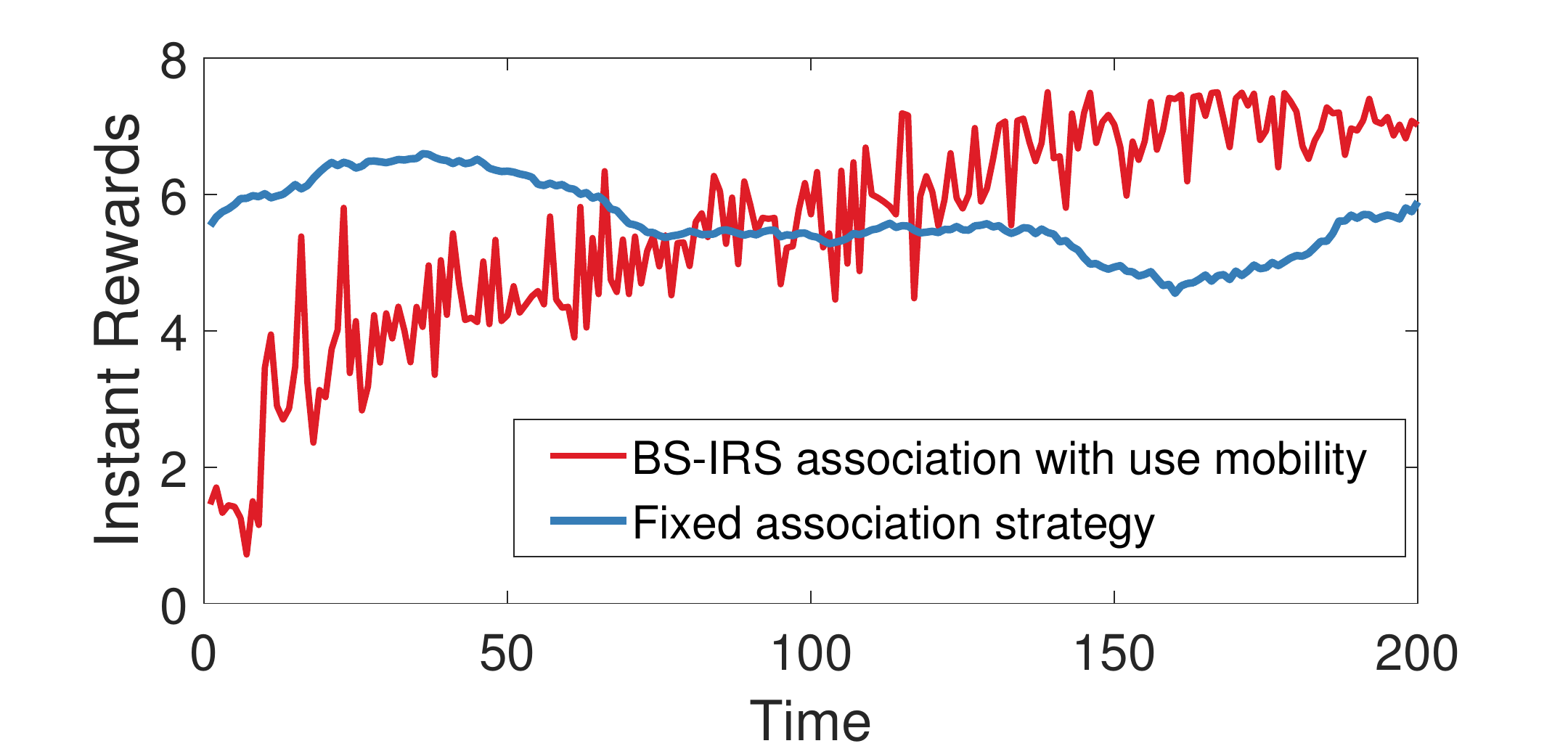}
	\caption{Instant reward (dB) as a function of time under different association strategies}
	\label{fig5}
\end{figure}

\section{Conclusion}
In this paper, we investigate the IRS-BS association problem in a mobile network consisting of multiple BSs serving a set of mobile users assisted by multiple IRSs.  In order to  maximize the long-term data communication performance for the associated users located in their service coverage areas, the BSs compete with each other for controlling the phase shift of a limited number of IRSs. A multi-agent reinforcement learning-based solution, named as MDLBI, is proposed to optimize the BS-IRS association and the  phase-shift of each IRS. The MDLBI achieves the maximum downlink communication sum rate without requiring any data exchange among BSs. Extensive simulations have been conducted to demonstrate that MDLBI achieves significant
performance improvement even when being implemented in large networking systems.

\section*{Acknowledgment}
This work was supported in part by the National Natural Science Foundation of China under Grants 62071193 and 61632019, the Key R \& D Program of Hubei Province of China under Grant 2020BAA002, China Postdoctoral Science Foundation under Grant 2020M672357.
\vspace{12pt}

\bibliography{ICC2021}
\bibliographystyle{IEEEtr}

\end{document}